\documentclass[fleqn,usenatbib]{mnras}
\usepackage[T1]{fontenc}
\usepackage{ae,aecompl}
\usepackage{graphicx}	
\usepackage{amsmath}	
\usepackage{amssymb}	
\usepackage{color}      

\newcommand{\grale}{\textsc{Grale}}
\newcommand{\lenstool}{\textsc{Lenstool}}
\newcommand{\glafic}{\textsc{Glafic}}
\newcommand{\arcsecond}{$^{\prime\prime}$}
\newcommand{\degree}{^\circ}
\newcommand{\simgt}{\hbox{\,\rlap{\raise 0.425ex\hbox{$>$}}\lower 0.65ex\hbox{$\sim$}\,}}
\newcommand{\simlt}{\hbox{\,\rlap{\raise 0.425ex\hbox{$<$}}\lower 0.65ex\hbox{$\sim$}\,}}

\title[Free-form \grale~reconstruction of Abell 2744]
{Free-form \grale~reconstruction of Abell 2744: robustness of uncertainties against changes in lensing data}

\author[K. Sebesta et al.]{Kevin Sebesta$^{1}$\thanks{Contact e-mail: \href{mailto:sebesta@physics.umn.edu}{sebesta@physics.umn.edu}},
Liliya L. R. Williams$^{1}$\thanks{Contact e-mail: \href{mailto:llrw@umn.edu}{llrw@umn.edu}},
Jori Liesenborgs$^{2}$,
Elinor Medezinski$^{3}$, and
\newauthor Nobuhiro Okabe$^{4,5,6}$\\
\\
$^{1}$School of Physics \& Astronomy, University of Minnesota, 116 Church Street SE, Minneapolis, MN 55455, USA\\
$^{2}$UHasselt - tUL, Expertisecentrum voor Digitale Media, Wetenschapspark 2, B-3590, Diepenbeek, Belgium\\
$^{3}$Department of Astrophysical Sciences, Princeton University, 4 Ivy Lane, Princeton, NJ 08544, USA\\
$^{4}${Department of Physical Science, Hiroshima University,1-3-1 Kagamiyama, Higashi-Hiroshima, Hiroshima 739-8526, Japan}\\
$^{5}${Hiroshima Astrophysical Science Center, Hiroshima University, 1-3-1 Kagamiyama, Higashi-Hiroshima, Hiroshima 739-8526, Japan}\\
$^{6}${Core Research for Energetic Universe, Hiroshima University, 1-3-1, Kagamiyama, Higashi-Hiroshima, Hiroshima 739-8526, Japan}
}
\date{Accepted XXX. Received YYY; in original form ZZZ}

\pubyear{2018}

\begin{document}
\label{firstpage}
\pagerange{\pageref{firstpage}--\pageref{lastpage}}
\maketitle

\begin{abstract}
Abell 2744, a massive Hubble Frontier Fields merging galaxy cluster with many multiple images in the core has been the subject of many lens inversions using different methods. While most existing studies compare various inversion methods, we focus on a comparison of reconstructions that use different input lensing data. Since the quantity and quality of lensing data is constantly improving, it makes sense to ask if the estimated uncertainties are robust against changes in the data. We address this question using free-form \grale, which takes only image information as input, and nothing pertaining to cluster galaxies. We reconstruct Abell 2744 using two sets of strong lensing data from the Hubble Frontier Fields community. Our first and second reconstructions use 55 and 91 images, respectively, and only 10 of the 91 images have the same positions and redshifts as in the first reconstruction. Comparison of the two mass maps shows that \grale~uncertainties are robust against these changes, as well as small modifications in the inversion routine. Additionally, applying the methods used in Sebesta et al. (2016) for MACS J0416, we conclude that, in a statistical sense, light follows mass in Abell 2744, with brighter galaxies clustering stronger with the recovered mass than the fainter ones. We also show that the faintest galaxies are anti-correlated with mass, which is likely the result of light contamination from bright galaxies, and lensing magnification bias acting on galaxies background to the cluster.
\end{abstract}



\section{Introduction}
\label{sec:intro}

The Hubble Frontier Fields program (HFF, PI: J. Lotz) is an unprecedented effort to observe and analyze six massive, merging galaxy clusters and neighboring parallel fields, over three years, $2013-2016$. Using about 850 orbits of Director's Discretionary time, HST made the deepest observations of clusters to date. The aim of the program was to use the magnification power of gravitational lensing by galaxy clusters to study high redshift background galaxies, as well as the clusters themselves \citep{lot17}. 

Over the span of the project, lens modelers used newly obtained HST data to reconstruct the six cluster mass distributions using methods that range from parametric (or simply parametrized), to hybrid, to free-form. The key difference between the parametric and hybrid vs. the free-form methods is that the former use information on cluster member galaxies and lensed images to solve for the cluster mass, whereas the latter rely solely on the images. Since the methods are quite different, and because even with $\mathcal{O}(100)$ lensed images per cluster the mass distribution is not completely constrained, employing a range of lens inversion techniques leads to a more thorough exploration of the possible mass distributions. In order to fully realize the science goals of the HFF program, it is important to compare the various lens inversion techniques, and to assess the systematic uncertainties in the reconstructed cluster properties, like the mass distribution and magnification power. Because of the collaborative effort between the teams of lens modelers, several studies carried out direct comparisons of the different methods \citep{pri17,men17,rem18}. Others have examined the effect of systematic uncertainties in mass reconstruction on the recovered luminosity function parameters of distant background galaxies \citep{bou17,ate18}, while still others looked into the systematics of using limited source redshift information when reconstructing synthetic clusters \citep{joh16}.

The first goal of the present work is also to address systematic uncertainties, but for a single method only, free-form \grale. Specifically, we evaluate the reliability, or robustness of \grale-derived uncertainties by comparing two reconstructions of the same HFF cluster, Abell 2744, that differ in input image properties, as well as certain lens inversion code parameters.  Our first reconstruction described in this paper, HFFv3, was done in the second year of the HFF project, using an image list with a combination of spectroscopic and photometric redshifts from \cite{ric14,joh14,wan15,jau15}. The following year, our second reconstruction, HFFv4, was completed with an all spectroscopic image list from \cite{mah18}. Both reconstructions use strong lensing data, consisting of positions of multiple images from background sources.

Since \grale~is fully free-form, and uses no information pertaining to the distribution of light from the cluster or its galaxies, it serves as an excellent method to accomplish our second goal, to investigate if there are any significant unexpected differences between the reconstructed mass and the observed light distributions. Following our previous report of a different HFF cluster, MACS J0416 \citep{seb16} we use mass-light centroid offsets, and galaxy-mass correlation function as our metrics to accomplish the second goal. 

Abell 2744 is a massive galaxy cluster with a right ascension of 00\textsuperscript{h}:14\textsuperscript{m}:19\textsuperscript{s} and declination of $-$30$\degree$:22$^{\prime}$:15$^{\prime\prime}$, and a redshift of $0.308$ \citep{man12}. The cluster is in an ongoing merger, first evidenced from a radio halo \citep{gio99,gov01a,gov01b}. The merger scenarios for Abell 2744 have been a popular topic of debate for the last 20 years \citep{owe11,mer11,med16,jau16}. Inversion methods have utilized many multiply-imaged systems \citep{zit14,joh14,ish15,jau15,kaw16,kaw18,mah18}, cluster-wide weak lensing distortions \citep{cyp04,med16}, or both \citep{mer11,lam14,wan15,jau16} to study the mass distribution at different scales. 

In Section \ref{sec:lit}~we give an overview of the previous mass models of Abell 2744. In Sec.~\ref{sec:grale} we outline our reconstruction technique and the inputs it uses. Section \ref{sec:results}~presents the comparison between our two reconstructions. In Section \ref{sec:xray}~we discuss the influence of mass outside the region of images on our results, and in Section \ref{sec:ltm}~we examine how well light traces mass in this cluster. Finally, Section \ref{sec:conclusion}~presents a summary of our findings. We use a flat $\Lambda CDM$ cosmology, with $\Omega_m = 0.3$ and $h = 0.7$ which results in a scale of $4.536$ kpc/\arcsecond~at the cluster redshift of $z=0.308$. If $\frac{D_{ls}}{D_s}$ is set to 1, the critical surface density for Abell 2744 is $\Sigma_{cr} = 0.3719$ g/cm$^2$.

\section{Existing Mass Models of Abell 2744}
\label{sec:lit}

  The first lensing mass estimates of Abell 2744 (also known as AC 118) were published by \cite{sma97} and \cite{all98}. \cite{sma97} constructed a weak lensing shear map for Abell 2744 from a catalogue of faint background objects built from HST data. Using this shear map and the singular isothermal sphere (SIS) profile to model the cluster, the cluster mass around the central core region, $r<400 h^{-1}$ kpc, was calculated to be $M = 1.85 \pm 0.32~10^{14} h^{-1} M_{\odot}$. The authors note that the configuration of the shear map reveals that the cluster contains two prominent mass clumps. \cite{all98} used strong lensing data from previous literature to calculate the mass of Abell 2744. Using an arc at $z = 1.00$, the mass of a circular distribution was calculated to be $M = 5.68~10^{13} h^{-1} M_{\odot}$.

\cite{cyp04} reconstructed the mass of Abell 2744, and $23$ other clusters, using the \textsc{LENSENT} software for weak lensing. Catalogues of galaxies within each cluster field were built, and galaxy ellipticity parameters were extracted using a Bayesian technique. Background galaxies were chosen based on their location behind the cluster and light properties, leading to accurate shape measurements. The total mass distribution was calculated using LENSENT, with a maximum entropy method, starting from the reduced shear field. After smoothing, two mass models, SIS and singular isothermal ellipsoid (SIE), were used to fit the cluster through minimization of $\chi^2$. Discrepancy between velocity dispersion calculated from weak-lensing data and that calculated from the dynamical study of \cite{gir01}, along with A2744 having the second highest cluster luminosity of the sample, further supported the idea that A2744 is the result of two merging clusters.

\cite{mer11} carried out the first comprehensive strong lensing and weak lensing mass reconstruction of Abell 2744. A parametric method was used to identify 34 multiple images from 11 background sources. Weak lensing analysis was done by combining three image sets (\textit{HST/ACS}, \textit{VLT/FORS1}, and \text{Subaru/SuprimeCam}) to get ellipticity measurements of background galaxies, and a final shear map. A total $\chi^2$ to be minimized consisted of a strong lensing term (dependent upon the lensing Jacobian), a weak lensing term (dependent upon the complex reduced shear), and a regularization term. The resulting mass, estimated as a function of radius, was $M (r < 1.3 Mpc) = 1.8 \pm 0.4~10^{15} M_{\odot}$, and $M (r < 250\,{\rm kpc}) = 2.24 \pm 0.55~10^{14} M_{\odot}$. The joint X-ray and lensing analysis revealed four main mass components, and provided clarity into the merging details of Abell 2744.

The CATS collaboration produced a mass model of Abell 2744 each year of the HFF program \citetext{v1: \citealp{ric14}, v2: \citealp{jau15}, v3: \citealp{jau16}, v4: \citealp{mah18}}. The algorithm \lenstool~was used to model cluster mass distributions from strong and weak lensing data. They represent the cluster's mass distribution by a few large cluster-scale haloes and galaxy-scale haloes. A Markov Chain Monte Carlo (MCMC) sampler is used to probe the parameter space and find the best fitting solution to the lens equation \citep{jul07}. \cite{ric14} used pre-HFF data of 55 images from 18 sources along with weak lensing data to obtain a best fitting model with lens-plane RMS of $1.26$\arcsecond. \cite{jau15} used only strong lensing data and two cluster-scale haloes to model the core. The best fitting model used 154 images from 54 sources and produced a lens-plane RMS of $0.79$\arcsecond. \cite{jau16} used new spectroscopic redshifts from \cite{wan15} to model Abell 2744 with a combined strong and weak lensing data set. Using 113 images from 39 different background sources, their best fit model produced a lens-plane RMS of $0.70$\arcsecond. \cite{mah18} performed an analysis of MUSE observations to extract new spectroscopic redshifts for objects in Abell 2744. An updated strong and weak lensing mass reconstruction obtained a lens-plane RMS of $0.67$\arcsecond~from 188 images and 60 systems.

\cite{joh14} also used the parametric method~\lenstool~lens inversion algorithm to model galaxy clusters using strong lensing data. A total of 47 images from 15 sources were used with redshifts either computed from Bayesian Photometric Redshifts (BPZ), measured from spectroscopy, or obtained from \cite{ric14}. Cluster member galaxies and 5, including 2 in the core, cluster-scale haloes were modelled with the pseudo-isothermal elliptical mass distributions (PIEMD). The iteration process uses minimization in the source plane, which is easier and computationally cheaper, but concludes with a minimization in the image plane. The resulting mass model with a lens-plane RMS of $0.40$\arcsecond~produced a central mass of $M(r < 250$~kpc$) = 2.43^{+0.04}_{-0.07}~10^{14} M_{\odot}$, comparable to that of \cite{mer11}.

\cite{zit14} used light traces mass (LTM) lensing model to reconstruct the mass distribution of Abell 2744. Their models assume that both the baryonic and dark matter components follow the light distribution, but to different degrees. A MCMC was used to find the best fit solution to the cluster's mass distribution. The best fitting model produced a lens-plane RMS of $1.3$\arcsecond. \cite{zit14} used another version of the algorithm, with elliptical NFW distributions, to check image positions predicted by LTM. This allowed them to find one of the most distant multiply imaged galaxies, at $z \approx 10$.

\cite{lam14} utilized a free-form lens reconstruction method, known as \textsc{WSLAP+}, to determine the cluster mass. The algorithm finds solutions for the surface mass density and positions of background sources in the source plane, by solving a set of linear equations. In addition to the free-form part to represent the cluster, parametric forms for cluster member galaxies were used. Model and photometric redshifts were used to identify several new image systems and correct several previously known systems, bringing the total image number to 65 from 21 background sources. The resulting mass model highlighted two major cluster components and significant excess mass near a large region of X-ray emission. The best mass model produced a lens-plane RMS of $1.25$\arcsecond.

\cite{wan15} used a free-form lens reconstruction method, called \textsc{SWUnited}, to reconstruct Abell 2744 using strong and weak lensing data, over an adaptive grid. This lens inversion method differs from most other methods in that it reconstructs lensing potential, instead of projected mass. Using multiply imaged sources and ellipticity of background galaxies as input, the algorithm minimizes $ \chi^2$ and converges to a final solution of the potential. Version v1 used pre-HFF data, including 44 images from 11 background sources. Using new spectroscopic Grism Lens-Amplified Survey from Space (GLASS) data, their selection algorithm found 72 images from 25 distinct sources. Their v2 mass model found two main mass peaks in the core and a lesser third peak north of the center.

\cite{med16} performed a weak lensing analysis of Abell 2744 using newly obtained imaging from Subaru/Suprime-Cam. The reduced shear from background galaxies was used to construct the mass distribution out to 5 Mpc, the largest extent probed by lensing in this cluster. Four substructures were found and modelled with NFW and truncated NFW profiles. Results suggested an alternative merging scenario compared to \cite{mer11}: the northwestern mass clump is the result of a smaller merger from two haloes that are falling into the main core of the cluster, instead of being produced from a `slingshot' effect of the main merger activity, as suggested by \cite{mer11}. 

\cite{jau16} also carried out a weak lensing reconstruction of the $\sim 1.7$Mpc extended area of the cluster. In addition to the core region, they identify seven other WL mass peaks. Four of these were also identified by either \cite{mer11} or \cite{med16}, or both, though the mass estimates of the latter two studies tend to be lower than those of the former. \cite{jau16} attribute this to the fact that \cite{med16} use only weak lensing data, and that the strong lensing images used by \cite{mer11} did not have spectroscopic redshifts, making mass normalization difficult.

The \glafic~team used a parametric reconstruction method applied to strong lensing data. The mass distribution of the cluster was modeled with cluster-scale NFW haloes, pseudo-Jaffe ellipsoids for cluster member galaxies, and external perturbations. A best fit model was found using a downhill-simplex algorithm to minimize a $\chi^2$. Errors of the mass models were derived through MCMC. \cite{ish15} used pre-HFF image data (67 images from 24 background sources) to model Abell 2744 and find faint galaxies at high redshifts. \cite{kaw16} used 111 images from 37 sources to produce their v3 map which had a lens-plane RMS of $0.37$\arcsecond. \cite{kaw18} increased the total number of images used to 132 from 45 sources for their v4 map, which produced a lens-plane RMS of $0.42$\arcsecond.

More gravitational lensing reconstructions of Abell 2744 can be found on the MAST website. \footnote{https://archive.stsci.edu/prepds/frontier/lensmodels/}

\section{Method and input}
\label{sec:grale}

\subsection{{\Large {\grale}}: free-form lens inversion method}

In this paper we use \grale~to produce mass maps of Abell 2744. \grale~is a free-form lens reconstruction method based on a genetic algorithm that finds solutions of the mass distribution of a lens. While parametric methods of lens reconstruction use many inputs related to the cluster galaxies, underlying cluster-wide dark matter haloes, and properties of lensed images, \grale~uses only the properties---positions and redshifts---of multiple images of background sources to derive the mass distributions. \grale~can model images as either point like, as was done in the analysis of HFF clusters MACS J0416 \citep{seb16} and MACS J1149 \citep{wil19}, or extended, as in the case of MACS J0717 \citep{wil18}. 

A \grale~run is initialized with a coarse uniform grid of fixed width cells, with each cell occupied by a projected Plummer sphere mass distribution, used as the basis set. The genetic algorithm then searches for the best solution of the mass distribution, in successive steps, according to a fitness value with two components, to be described later. At each step, the genetic algorithm refines the original grid, subdividing it into smaller grid cells where mass density is high, and placing a Plummer sphere in each grid, with its width matching the size of the grid cell, and its mass determined by the constraints provided by the lensed images. The genetic algorithm uses mutation and reproduction to breed new solutions of the mass distribution. This process is repeated until a final solution is found.

The fitness measure of a genetic algorithm assesses how well a mass distribution satisfies the lens equation, and is therefore used to select the best solution for any given \grale~run. We use the same fitness measures as those used in \cite{seb16}, but because we use extended images instead of points images, the implementation of the fitness measures is slightly different. The first fitness measure is based on the fractional overlap of the observed extended images, when backprojected to the source plane. Using fractional overlap, instead of actual overlap in terms of arcseconds, guards against solutions that overfocus the images. Mass reconstructions that lead to backprojected images of the same source having a higher fractional overlap, are assigned better fitness values.  The other fitness measure makes use of the null space, or the area in the image plane that contains no images. For extended images, the area occupied by observed images is cut out from the null space. Furthermore, holes in the null space are made for regions where one suspects additional images, based on lensing theory. Each source has its own null space. Mass distributions that predict images in the null space are assigned worse fitness values. 

\grale~uses a multi-objective genetic algorithm to optimize these two fitness measures at the same time, without requiring any regularization. The method is described in more detail in \cite{lie06, lie07, moh14, seb16, wil18}. Next, we describe two sets of Abell 2744 reconstructions,  HFFv3 and HFFv4, carried out using multiple image data shared by the HFF community.

\subsection{HFFv3}
\label{sec:summer15}

We generated 40 \grale~reconstructions based on 55 multiply lensed images from 18 different sources for HFFv3. The image list was compiled by the HFF community data presented in \cite{ric14,joh14,wan15,jau15}. Six sources had spectroscopic redshifts from \cite{ric14,joh14,wan15}, and the rest of the sources we used had their redshifts fixed to the model redshifts from \cite{jau15}. Each reconstruction was the result of finding the best solution of nine successive grid refinements. We increased the number of Plummer spheres in each subsequent grid by approximately 200. The number of Plummers in the first and last grid were chosen by the genetic algorithm from ranges of 300-400 and 1700-1800, respectively. The final solution for each reconstruction was chosen based on the two fitness values, described above, for all nine grids. Each reconstruction is started with its own random seed, and is somewhat different from the rest because of the very large dimensionality of the parameter space that the genetic algorithm searches. An average is taken of these 40 reconstructions, which highlights the common features in the mass distribution, and suppresses rare, one-off features.

\subsection{HFFv4}
\label{sec:fall16}

For HFFv4, we produced 40 \grale~reconstructions using 91 multiply lensed images from 29 different sources. All 91 images used updated spectroscopic redshifts from \cite{mah18}. Note, for a system of images from the same source, \cite{mah18} may not have found a spectroscopic redshift for every image, but we chose to use the images without spectroscopic redshifts, assigning the redshift of the system. There is some overlap between the image lists for our HFFv3 and HFFv4. There are 29 images, from 10 sources, that have the same positions for both HFF versions. Of these, 10 images (from 3 sources) have a difference in redshift $\lvert{z_{v3}-z_{v4}} \rvert < 0.1$, whereas the rest have a larger difference in redshift up to $\lvert{z_{v3}-z_{v4}} \rvert = 1.42$. We used only images with spectroscopic redshifts because they typically have a higher degree of accuracy than redshifts predicted by lens inversion models.

The process for producing mass distribution solutions for HFFv4 is similar to that for HFFv3 described above, with one exception. The number of Plummer spheres in each grid refinement no longer increased linearly with grid number. Instead, the number of Plummer spheres increased by, on average, 175, starting from the first to seventh grid, ending with a maximum range of 1100-1200. The eighth and ninth grid had approximately 150 fewer Plummer spheres than the preceding grid. This procedure, similar to what was used in \cite{moh14}, and different from that used for Abell 2744 in HFFv3, was implemented to test if reducing the number of Plummer spheres, and hence parameters, will retain the goodness of fit achieved in the seventh grid. The best mass map for each run was selected according to the two fitness values, from the nine grids. The eighth and ninth grids were the most frequently preferred by \grale, showing that \grale~does not always converge to the grid with the highest number of Plummer spheres. As for HFFv3, an average of the 40 individual reconstructions was calculated as the best solution for HFFv4. This HFFv4 map has been used in \citep{mon18}.

\section{Results of Lens Reconstruction}
\label{sec:results}

\begin{figure*}                
  \includegraphics[width=\textwidth]{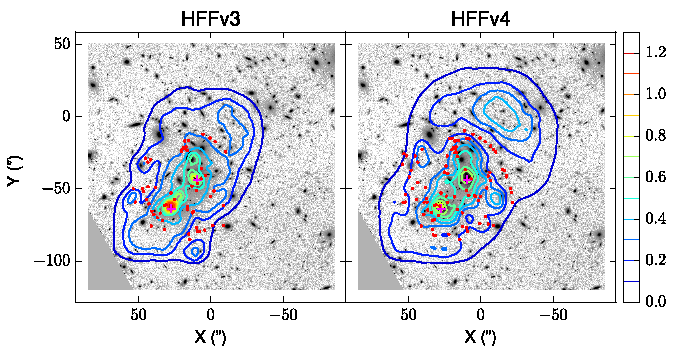}
  \caption{Mass contours of averaged mass map of Abell 2744 overlaid on a HST total (F435W, F606W, and F814W combined) image. {\it Left:} HFFv3; {\it Right:} HFFv4. The lines represent mass contours in units of g/cm$^2$, and red circles represent images used in the two reconstructions. The two BCGs are shown as magenta crosses.} \label{fig:avgmap}
\end{figure*}

 The average mass maps of the 40 individual \grale~reconstructions for HFFv3 and HFFv4 are shown in left and right panels, respectively, of Fig.~\ref{fig:avgmap}. The contour lines represent the projected surface mass density, $\Sigma$, starting from 0.1 g/cm$^2$ and linearly spaced by 0.1 g/cm$^2$. The left (right) panel has twelve (eight) lines, with the last one at 1.2 g/cm$^2$ (0.8 g/cm$^2$). The red filled circles represent the images used as input for each version. The two brightest cluster galaxies (BCGs) are highlighted as magenta crosses. The elongated shape of Abell 2744 is clearly visible in both maps. Likewise, the two major cluster components that have been identified by previous reconstructions, are found by both maps. HFFv3 shows steeper mass peaks around the two BCGs, relative to HFFv4; we will return to this in Section~\ref{sec:ltm}. Both versions produced good lens-plane RMSs, $0.53$\arcsecond~and $0.87$\arcsecond~for HFFv3 and HFFv4, respectively. The fact that HFFv4, which has better quality data, has larger rms, is not necessarily unexpected. \cite{joh16} find that the lens-plane rms increases with the number of systems used in reconstruction (see their Fig.3, second panel in the bottom row). However, a more detailed comparison with that figure may not be applicable, as the assumptions of their study are quite different from ours.

The mass enclosed within 250 kpc of the cluster center is $M(r<250\,{\rm kpc})=2.25 \pm 0.06 \times 10^{14}M_\odot$ and $2.27 \pm 0.06 \times 10^{14}M_\odot$ for HFFv3 and HFFv4, respectively, compatible within $\sim 2\sigma$ with the masses found by \cite{mer11}, \cite{joh14} and \cite{wan15}; see Table~\ref{tab1}. The mass within the same radius found by \cite{jau15} and \cite{jau16} is $\sim 20\%$ larger than ours, and discrepant with ours at a high significance level. While the details of the reconstructed mass distribution can depend on the inversion method used, such a large difference in total enclosed mass is hard to attribute to different lens inversion methodologies. Since the image sets and their total number used by Jauzac et al. and here are not exactly the same, we speculate that the difference in total mass is due to the difference in positions or redshifts assumed for some of the less secure images, used in one or both of the reconstructions.

\begin{table*}
        \centering
        \caption{Comparison with other published results. The columns are: (1) model, (2) the name of the lens reconstruction method, (3) whether the reconstruction was based on strong lensing and/or weak lensing data, (4) the number of strong images (and corresponding sources) used in the reconstruction, (5) derived mass within 250 kpc.}
        \label{tab1}
        \begin{tabular}{lcccl} 
                \hline
                Models & Inversion method & SL and/or WL & SL images (sources) & $M(<250{\rm kpc}), ~10^{14}M_\odot$ \\
                \hline
                This work: HFFv3& \grale~    & SL    & 55 (18)  & $2.25\pm 0.06 $\\
                This work: HFFv4& \grale~    & SL    & 91 (29)  & $2.27\pm 0.06 $\\
                \cite{jau16}    & \lenstool~ & SL+WL & 113 (39) & $2.77\pm 0.01 $\\
                \cite{jau15}    & \lenstool~ & SL    & 154 (54) & $2.762\pm 0.006 $\\
                \cite{joh14}    & \lenstool~ & SL    & -- (15)  & $2.43^{\, +0.04}_{\,-0.07} $\\
                \cite{wan15}    & SWUnited   & SL+WL & 72 (25)  & $2.43^{\,+0.04}_{\,-0.03} $\\
                \cite{mer11}    & SAWlens    & SL+WL & 34 (11)  & $2.24\pm 0.55 $\\
                \cite{med16}    & ---        & WL    & ---      & $1.49\pm 0.35 $\\
                \hline
        \end{tabular}
\end{table*}

\begin{figure*}                
  \includegraphics[width=\textwidth]{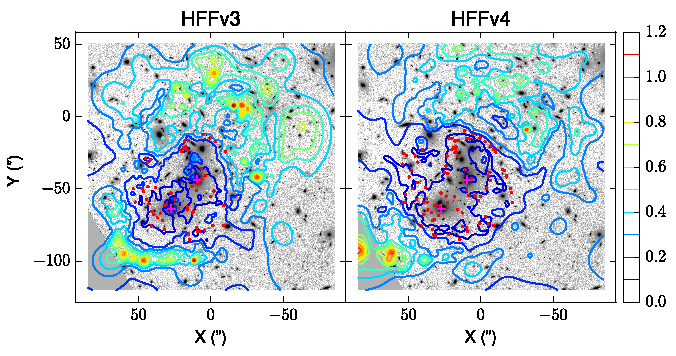}
  \caption{Contours of fractional uncertainty in surface mass density of Abell 2744, overlaid on the same HST image as in Fig.~\ref{fig:avgmap}, with same field of view. {\it Left:} HFFv3; {\it Right:} HFFv4. The contours are linearly spaced by 10\% and start at 10\% (dark blue line). The left panel has a total of ten contour lines and the right panels has a total of eleven contour lines. Images are highlighted as red circles, and the two BCGs are marked as magenta crosses.} \label{fig:fracmap}
\end{figure*}

 Next, we look at the fractional mass uncertainty in the HFFv3 and HFFv4 sets of 40 individual reconstructions. For a set of mass maps, the fractional uncertainty at any point in the lens plane is $\epsilon / \Sigma$ where $\epsilon$ is the root-mean-square deviation in the maps' $\Sigma$ values. Fig.~\ref{fig:fracmap} shows the fractional uncertainty map for HFFv3 (left panel) and HFFv4 (right panel).  Similar to MACS J0416 in \cite{seb16}, \grale~produces low fractional uncertainty within the cluster region where images lie. Most of the cluster's elongated shape, the central $\sim 20$\arcsecond$\times 60$\arcsecond, has errors below 10\% for HFFv4, and below 20\% for HFFv3. Circular regions of radius $r \approx 2.5$\arcsecond~around the two BCGs have errors of less than 10\%  in both versions. Outside the center of the cluster, where there are no images, the fractional uncertainty is $\simgt 30-50$\%, with no clear pattern; \grale~is not constrained here and is therefore less reliable in reconstructing the mass distribution.

\begin{figure*}                
  \includegraphics[width=\textwidth]{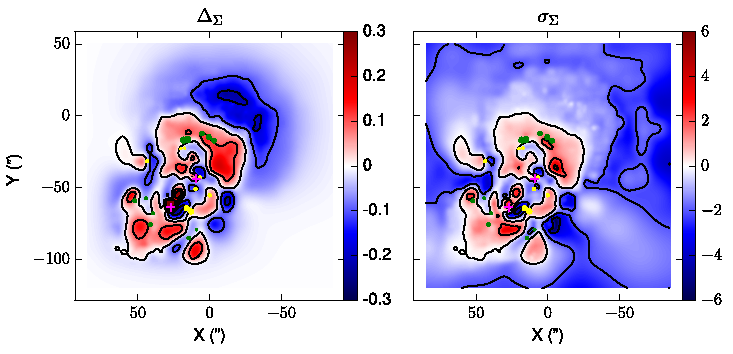}
  \caption{\textit{Left:} Difference map of the two mass reconstructions of Fig.~\ref{fig:avgmap}, $\Delta_{\Sigma} = \Sigma_{v3} - \Sigma_{v4}$, where v3 and v4 are for HFFv3 and HFFv4 respectively. The solid contour lines are at $\Delta_{\Sigma} =0,\pm 0.1,\pm 0.2$.  \textit{Right:} Map of the statistical significance of the difference map. The solid contour lines are at $\sigma_{\Sigma} =0,\pm 2,\pm 4$. The yellow and green circles represent images that have a difference in redshift of $({z_{v3}-z_{v4}}) > 0.1$ and  $({z_{v3}-z_{v4}}) < -0.1$, respectively. The size of the circle corresponds to the magnitude of difference in redshift.  Both panels show the same region as in Fig.~\ref{fig:avgmap} and Fig.~\ref{fig:fracmap}.} \label{fig:diffmap}
\end{figure*}

In addition to analyzing the results of our reconstructions, we also present the comparison of HFFv3 and HFFv4. Prompted by the Hubble Frontier Fields project, several papers have examined the statistical and systematic uncertainties in lens reconstruction methods \citep{men17,pri17,rem18,chi18}. Most of these analyses compared different lens inversion methods, while keeping the input data as similar as possible. Our approach is the opposite: we use (nearly) the same lens inversion method, but different input. We compare two reconstructions of Abell 2744 using \grale, that are based on different data sets, in terms of images used, image positions, and source redshifts. Only 10 out of 91 images used in HFFv4 are the same as in HFFv3 in terms of positions and redshifts. The two implementations of \grale~are also somewhat different, as described in Sections~\ref{sec:summer15} and \ref{sec:fall16}. This type of comparison is closest to what modelers and those reading their papers encounter: two lens reconstructions by the same group, done a year or two apart, where the second reconstruction was motivated by an improved and enlarged data set. In most realistic situations, the inversion code itself has also undergone some modifications.

The left panel of Fig.~\ref{fig:diffmap} shows the difference map of HFFv3 and HFFv4, $\Delta_{\Sigma} = \Sigma_{v3} - \Sigma_{v4}$.  A more meaningful understanding of the difference map can be gained by considering the significance, defined as $\sigma_{\Sigma} = \Delta_{\Sigma} / \sqrt{\epsilon_{v3}^2 + \epsilon_{v4}^2}$, which highlights regions of the lens plane where discrepancies in the mass distributions are statistically significant. The $\sigma_{\Sigma}$ map is shown in the right panel of Fig.~\ref{fig:diffmap}. 

Since \grale~relies solely on image data as input, it is instructive to see if there is a direct connection between changes in input data and changes in the resulting mass distribution. The circles in Fig.~\ref{fig:diffmap} represent images, used in both reconstructions, that have a significant change in input redshift, $\lvert{z_{v4}-z_{v3}} \rvert > 0.1$. Yellow circles are images with a positive difference (HFFv4 redshift is larger) and green circles are images with a negative difference (HFFv3 redshift is larger). The radius of the circle is scaled to the change in redshift, with the largest difference being $\lvert z_{v3}-z_{v4}\rvert = 1.42$. The distribution of these images reveals no clear pattern in the difference or significance map, implying that the input image data affects the mass distribution on most of the lens plane.

The steeper density profiles in the few arcseconds around the two BCGs in HFFv3 vs. HFFv4 (Fig.~\ref{fig:avgmap} and the left panel of Fig.~\ref{fig:diffmap}), has $\sim 2\sigma$ significance, in the right panel of Fig.~\ref{fig:diffmap}. Though the difference is not statistically significant, it is still interesting to understand the reason for it. The peaks in the difference and significance maps surrounding the Southern BCG are caused by the higher concentration of Plummer spheres in HFFv3 compared to HFFv4. Recall from Section \ref{sec:grale} that HFFv3 reconstructions were allowed to use more Plummer spheres as inputs than HFFv4. Combined with the fact that \grale~refines the grid based on the local mass density, which is high close to galaxies, and that HFFv3 was allowed to continue to refine the grid in the eighth and ninth solutions, resulted in the HFFv3 average map having roughly four times as many Plummer spheres within a $4$\arcsecond~radius around the Southern BCG compared to the same region in the HFFv4 average map. It is important to note that this higher concentration of mass, or higher steepness, is possible because there are no images within $\sim 5$\arcsecond~of the BCG center to act as model constraints. In regions with no constraints, the monopole degeneracy \citep{lie08} is free to redistribute the mass in any circularly symmetric, mass conserving fashion.

The differences between the two mass maps, expressed in units of \grale~derived errors (right panel of Fig.~\ref{fig:diffmap}) tells us how robust \grale~reconstructions are to changes in input lens data and \grale~implementation procedure, and hence how reliable \grale's estimate of uncertainties is. In the case of Abel 2744, some of the image data used in HFFv3 turned out to be erroneous, and were corrected in HFFv4. The area of regions with significant differences in mass distribution, $\sigma_{\Sigma}\ge 4$, is $75\,\sq$\arcsec, or about 1.5\% of the cluster area covered by multiple images. Since most of the mass maps of HFFv3 and HFFv4 are within each other's errors, \grale-derived uncertainties can be trusted.

It is instructive to compare this conclusion with that based on Abell 3827, which was reconstructed using HST data \citep{mas15}, and HST and ALMA data \citep{mas18}. The first reconstruction detected an offset, $\sim 1.6\,$kpc, between the light of one of the four cluster ellipticals and the nearby mass peak. The offset was seen in \grale~as well as \lenstool~mass maps, and was attributed to self-interacting dark matter. The second reconstruction, based on a somewhat different data set did not detect an offset, with either \grale~or \lenstool. Applying the $\sigma_\Sigma$ analysis to the two \grale~maps of Abell 3827, we see that within the image circle, $\sim 0.25\%$ of the area has $\sigma_\Sigma\ge 4$. The $\sim 3\arcsec\times 3\arcsec$ region around the elliptical in question has a range of $\sigma_\Sigma$ values, from $1-3.5$, but does not reach $4$. So our conclusion that \grale's $\sigma_\Sigma$ uncertainties are robust against reasonable changes in input multiple image data applies to Abell 3827 as well.

Somewhat different, but consistent findings about \grale-derived uncertainties are presented in \cite{rod15} Figure 6, where \grale's uncertainties on the magnification of Type Ia SN HFF14Tom cover nearly the full range of uncertainties of all other lens inversion methods, as well as the true magnification of the supernova. \cite{pri17} extended this type of magnification analysis to the whole strong lensing regions of MACS J0416 and Abell 2744, and, in general, confirmed the conclusion obtained from a single lens plane location of SN HFF14Tom.

\begin{figure*}                
  \includegraphics[width=\textwidth]{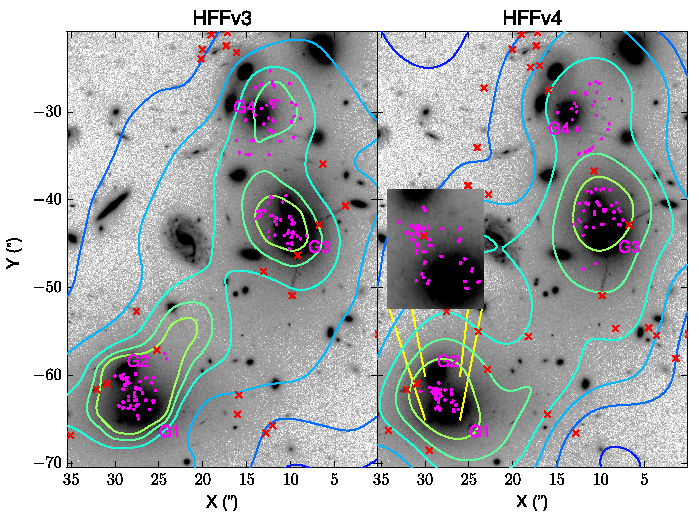}
  \caption{Zoomed in area of Fig.~\ref{fig:avgmap}, spanning $35.6$\arcsecond~by~$49.6$\arcsecond. {\it Left: } HFFv3; {\it Right:} HFFv4. Magenta points represent peaks of 40 mass reconstructions in the projected surface mass density around four main galaxies, labelled G1-G4. These points were slightly displaced from their positions by random amounts to avoid overlap and make all the points visible. (The points were obtained on a grid, so many were superimposed.) Red crosses are the images used for each reconstruction version. Contour lines match the scale from Fig.~\ref{fig:avgmap}, but only extend up to 0.7 g/cm$^2$. Higher density contour lines were left out for clarity. A zoomed in region around G1 and G2 of HFFv4 is provided in the right panel. It uses a lighter gray scale shading to highlight the region between the two galaxies.}\label{fig:pkmap}
\end{figure*}

\section{Influence of the mass to the North West of the main cluster}
\label{sec:xray}

\grale~reconstructions show a prominent broad peak North West of the main cluster, with HFFv4 showing a larger mass excess than HFFv3. It is centered at approximately ($-15$\arcsecond, $0$\arcsecond) in Fig.~\ref{fig:avgmap}, and about $50$\arcsecond~to the North West of the northern BCG. The peak is beyond the image region, and hence outside the region that \grale~can constrain well. The fractional uncertainty, $\simlt 50\%$, is high enough so that the detailed features of this mass clump cannot be taken too seriously, but the existence of extra mass in that direction with respect to the main cluster appears to be robust.

This mass clump is at the same location as the broad peak in the X-ray distribution of Abell 2744, seen, for example in the maps of X-ray luminosity contours presented in \cite{mer11}. \cite{owe11} and \cite{jau16} present an extensive description of the X-ray emitting gas based on the analysis of {\it Chandra} and {\it XMM-Newton} data, respectively. They identify several X-ray features, some of which are associated with the weak-lensing mass peaks. Neither of these studies estimate the mass of the gas responsible for the X-ray emission. However, despite the coincidence in the location of the peak, the mass in the X-ray emitting gas is unlikely to be large enough to cause the mass clump. \cite{gov01b} examined the X-ray emission of Abell 2744, as a part of a larger study. From their Fig. 14, the X-ray mass within $r\approx 100$ kpc of the X-ray peak is $M\approx 5\times 10^{11} M_{\odot}$. Within the same radius, \grale~finds $M_{v3} = 3.17 \pm 0.47 \times 10^{13}M_{\odot}$ and $M_{v4} = 4.88 \pm 0.61 \times 10^{13} M_{\odot}$, or about 100 times more than the mass in X-ray emitting gas. Assuming the X-ray gas mass to be an external point source, located at the peak of the X-ray emission, outside the main cluster, the amplitude of deflection angles at the North West and South East ends of the lensed image distribution are $\alpha \approx 0.16$\arcsecond~and $\alpha \approx 0.04$\arcsecond, respectively. Since these are smaller than the lens-plane RMS of any reconstruction, it is unlikely that either \grale~or any other lens inversion method would be able to discern the X-ray gas. 

Instead, the excess mass seen by \grale~provides external shear, which is due to the merging cluster, located in the same direction, about $150$\arcsecond~ North West of A2744. This cluster is seen directly in wider field weak lensing reconstructions \citep{mer11,wan15,jau16,med16}.

\begin{figure}                
  \includegraphics[width=\columnwidth]{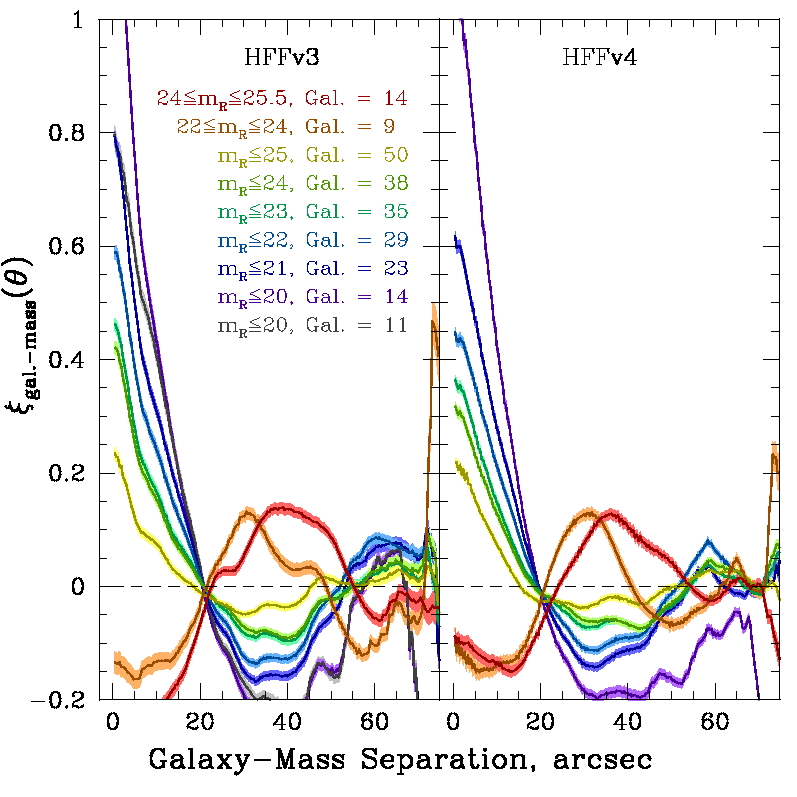}
  \caption{The left and right panels show normalized galaxy-mass correlation functions for R-band Subaru galaxies, for HFFv3 and HFFv4, obtained using all 40 individual \grale~maps, and 10 random galaxy realizations for each. These calculations were done in a circular region centered on the core of Abell 2744, and using eight galaxy magnitude cuts, as indicated in the figure. Each magnitude cut is color coded and lists the number of galaxies contained in that magnitude range. An additional correlation function (gray) is plotted for HFFv3, which is the same as the $m_R \le 20$, but with galaxies G1-G3 removed. Shaded color bands are 1$\sigma$ deviation of the mean error bars. (To obtain the RMS, one must multiply by $\sqrt{400}$. The typical one-sided RMS dispersions along the vertical axis are between 0.1 and 0.25 in both panels.)}\label{fig:corr_sub}
\end{figure}

\section{Comparing the distributions of light and mass in Abell 2744}
\label{sec:ltm}

In this section we look at how closely the distribution of mass follows the light distribution of cluster member galaxies. \grale~is well suited for this purpose, since no information about member galaxies is used as input in creating the mass maps. Only one other existing technique is similar to \grale~in this regard---\textsc{SWUnited} \citep{wan15}---all others use parametrized forms to represent mass in cluster galaxies. We utilize two types of analysis to quantify how closely cluster member galaxies of Abell 2744 follow the overall mass distribution.

\subsection{Local Mass Peaks near Massive Galaxies}
\label{sec:peaks}

Fig.~\ref{fig:pkmap} zooms in on the central region of Fig.~\ref{fig:avgmap}, containing the four central massive galaxies, labeled in magenta as G1-G4. According to the right panel of Fig.~\ref{fig:fracmap} this region has fractional uncertainty $\simlt 10\%$ for HFFv4 reconstruction, i.e. it is well constrained. For HFFv3 (left panel), the fractional uncertainty is larger, $\simlt 20-30\%$, but still low enough for consideration. As indicated by the mass density contour lines, both HFFv3 and HFFv4 pick out the two prominent mass concentrations, associated with G1-G2 and G3. These correspond to the centers of the cluster-scale dark matter haloes used in previous literature. Galaxy G4 is also identified by \grale, but many galaxies further away from the center are not; in several cases mass contour lines go right through fainter galaxies. This means that lensed images alone do not require mass concentrations at the locations of these galaxies. In parametric methods, whose reconstructed mass distributions would include these galaxies, it is the model priors based on the presence of light that are responsible for mass at these locations, with minimal input from multiple images.

In Fig.~\ref{fig:pkmap}, the contour lines represent projected density distribution of the average \grale~maps, and the magenta dots show the local mass density peaks, within a circle of radius 5\arcsecond~centered on the four galaxies, G1-G4, in each of the 40 individual reconstructions. The dispersion of these points on the lens plane gives an estimate of the uncertainty in the location of the mass peaks associated with the massive central galaxies. The clouds of magenta points are centered on the corresponding galaxies, consistent with there being no detectable offsets between the center of light and the center of mass, on scales larger than $\sim 3$\arcsecond, or $\sim 14$ kpc. This is consistent with the findings of \cite{mas18} in the case of Abell 3827.

The distribution of magenta points around G1 and G2 galaxies is more compact in HFFv4 compared to HFFv3, and requires some explanation. The inset in the right panel of Fig.~\ref{fig:pkmap} shows a zoomed in version of this region. We believe that the reason for this compact distribution is a single observed lensed image, located between G1 and G2. This image is a part of a five-image configuration. In the absence of any information about the light distribution, \grale~is interpreting this system as a regular ``quad'' with a central image. However, this is most likely a misinterpretation.  Because of the presence of adjacent G1 and G2 near that image, instead of a single mass peak, the actual image configuration is probably more complex, and the image multiplicity is seven. The observed central image is likely a saddle point in the arrival time surface, with two galaxies producing nearby maxima in the arrival time. (This triple is surrounded by the 4 images of a quad.) But \grale~prefers the simpler five-image solution, where the central image is a maximum in the arrival time surface. There are no other images within $\sim 2-3$\arcsecond~to help \grale~with the interpretation.

\subsection{Mass-Galaxy Correlation Function}
\label{sec:corrfunc}

The second technique we will use to investigate how well light traces mass is the 2D correlation function between cluster member galaxies and average mass maps, $\xi_{gm}(\theta)$. The probability of finding a second galaxy in a small area $dS$, at distance $\theta$ away from a galaxy is $dP=n(1+\xi_{gm})dS$, where $n$ is the average surface number density of galaxies within that cluster. The metric $\xi_{gm}(\theta)$ will help us determine if total mass, which is mostly dark matter, clusters with visible galaxies in Abell 2744. We use the same estimator as in \cite{seb16}, ~$\xi_{gm}(\theta)=\frac{D_gD_m(\theta)}{\langle R_gD_m\rangle(\theta)}-1$, where $D_gD_m$ is the number of galaxy-mass pixel pairs, and $\langle R_gD_m\rangle$ is the number of random galaxy-mass pixel pairs. To calculate the mean and uncertainties we used all 40 individual \grale~maps, and 10 random galaxy realizations for each, for a total of 400 realizations. (Here, $D$ stands for direct, and $R$ stands for random.)

We use the galaxy catalogue of \cite{med16} based on the Subaru data which encompasses the total field of view for Abell 2744.  We use Subaru, instead of HST cataloque for a number of reasons.  Subaru catalogue has magnitudes for all the HST galaxies, including those that are too bright for HST. The Subaru filters we use, Z and R, are closely matched by the HST filters, F814W and F606W, respectively. Furthermore, for the analysis described later in this section, we need galaxy number counts in a field larger than that of the HST, to avoid lensing effects of the cluster. Finally, our earlier paper on another HFF cluster, MACS J0416 \citep{seb16} also used Subaru galaxies.

We restrict our galaxy-mass correlation calculation to the core of Abell 2744, where strong lensing features are present. We use a circular region centered on $\alpha=$00\textsuperscript{h}:14\textsuperscript{m}:21.278\textsuperscript{s}, $\delta=-$30$\degree$:24$^{\prime}$:04.67$^{\prime\prime}$, with radius $r=39$\arcsecond~and a bin size of $0.39$\arcsecond. This region encompasses all the images of HFFv3 and most of the images of HFFv4, along with the two BCGs in the core of Abell 2744. Our choice of Subaru galaxies requires masking to deal with the light of bright galaxies that may block nearby fainter galaxies. When calculating the galaxy-mass correlation functions, we mask regions around the 10 brightest galaxies with a circle of variable radius ($4$-$11$\arcsecond), which scales with the galaxy brightness profile in Subaru images. 

The correlation functions between Subaru R-band galaxies and averaged HFFv3 and HFFv4 mass maps are shown in the two panels of Fig.~\ref{fig:corr_sub}. We repeated all the analyses with Subaru Z-band, but since the results are very similar we do not show the correlation functions for that band. We chose eight magnitude cuts to investigate the behavior of \grale~reconstructions against Subaru galaxies. The numbers of galaxies used in each magnitude cut are listed in the figure. The error bars represent the deviation of the mean, i.e., how much the mean of the 400 realizations can vary. (To get the RMS between the 400 realizations, one has to multiply by $\sqrt{400}$, resulting in typical one-sided RMS values of $0.1-0.25$.) The deviation of the mean is the more appropriate quantity since we are interested in the mean statistical trends, and not how much individual \grale~realizations can vary between each other.

The galaxy-mass correlation functions for both versions of \grale, depicted in Fig.~\ref{fig:corr_sub}, show similar behavior, clearly indicating that the clustering amplitude of the reconstructed total mass scales with the brightness of the galaxies. However, comparing HFFv3 and HFFv4 within each galaxy color shows that $\xi_{gm}(\theta)$ of HFFv3 has a higher amplitude than that of HFFv4 for bright magnitude cuts. This is a direct consequence of the mass distribution around galaxies being less peaked in HFFv4 compared to HFFv3, already seen in Fig.~\ref{fig:avgmap}. In addition to being more peaked, $\xi_{gm}(\theta)$ of HFFv3 show small breaks, or elbows, around $\theta\approx 6-7$\arcsecond, which can be attributed to the very central mass peaks in HFFv3 being more concentrated than in HFFv4. To test if the most massive galaxies are responsible for the elbows, we removed the galaxies G1-G3 from the $m_R \le 20$ magnitude bin and computed the normalized correlation function for HFFv3, shown as a gray band in the left-most panel of Fig.~\ref{fig:corr_sub}. This correlation function has smaller initial amplitude, close to the one for the $m_R \le 21$ magnitude cut, then falls off more gradually and connects to the $m_R \le 20$ correlation function around $12$\arcsecond~separation. Since the elbow persists, we conclude that the elbows are due to a wide range of galaxies, and are likely the result of the differences in \grale's resolution in HFFv3 vs. HFFv4, discussed in Section~\ref{sec:results}. 

The trend of brighter galaxies clustering stronger with mass matches the results of our previous study of \grale's reconstruction of MACSJ0416, another HFF galaxy cluster. The standard biasing scenario of galaxy formation states that galaxies are biased tracers of the underlying mass distribution and \grale's results are further validations of this. Note that even though we did not select galaxies based on cluster membership, at brighter magnitudes most, if not all galaxies are cluster members, as can be seen in Fig.~\ref{fig:m_z}.

The galaxy-mass correlation function is a powerful tool to investigate how well light traces mass in galaxy clusters, but caution needs to be exercised when working with fainter galaxies. Whereas bright galaxies are most likely cluster members, some faint galaxies can be background galaxies that underwent lensing magnification, or did not make it into the sample because of the light contamination from nearby bright galaxies.

Therefore, correlation function between mass and faint galaxies can be affected by both of these effects. The lens magnification bias is the result of two effects. The area behind the lens is enlarged, thereby diluting the number density of background galaxies. The second effect, competing against the first, is that background galaxies are magnified to appear brighter than unlensed galaxies at the same redshifts. The slope of the unlensed number counts of background galaxies determines which of the two effects dominates. The two effects cancel each other when the slope of $d\log(n[f])/d\log(f)=1$, or $d\log(n[m])/dm=0.25$, if magnitudes are used.

\begin{figure}                
  \includegraphics[width=\columnwidth]{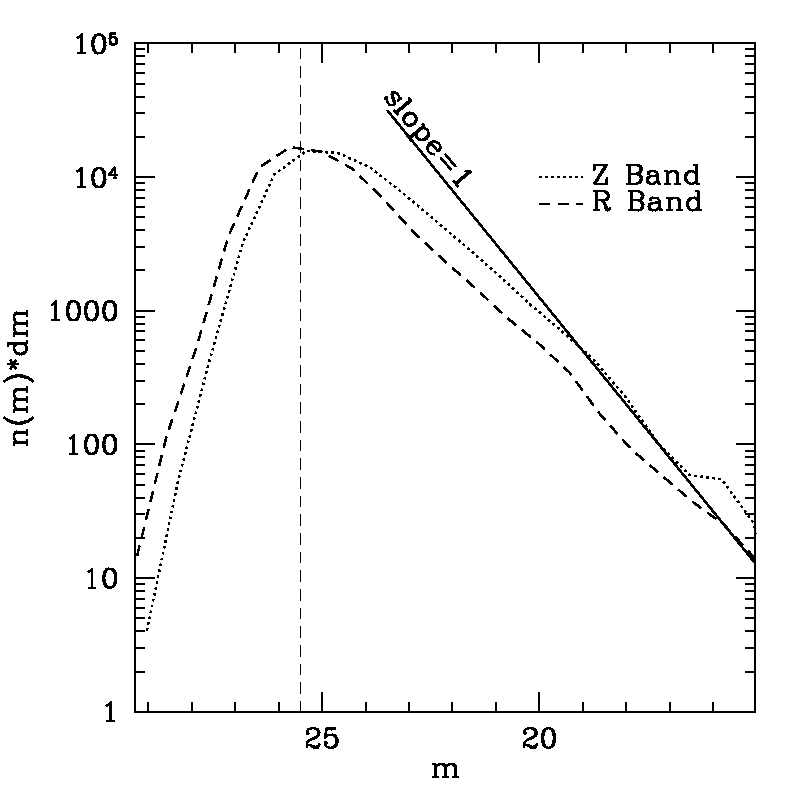}
  \caption{Differential number galaxy counts versus magnitude of Subaru R-band and Z-band galaxies. Because of low number counts in the cluster region, the field of view used is the entire Subaru region. A dashed vertical line indicates where the catalogue suffers from incompleteness. A solid line with slope, $d\log(n[f])/d\log(f)=1$, is provided for reference.}\label{fig:numcount}
\end{figure}
 
The differential galaxy counts of the entire Subaru field as a function of galactic magnitude for Subaru R-band, Z-band are plotted in Fig.~\ref{fig:numcount}. A line of slope $d\log(n[f])/d\log(f)=1$ is provided for reference. Since most of the galaxies are not in the direction of the central region of Abell 2744, these counts are, effectively, unlensed. We do not consider any galaxies fainter than $m \approx 25.5$, because the counts begin to flatten off and suffer from incompleteness.

At magnitude $m \approx 19$ there is a slight break in the number counts of both Z-band and R-band. Since the slope becomes shallower than $d\log(n[f])/d\log(f)=1$ at fainter magnitudes, the net effect of the magnification bias is to decrease the galaxy counts behind a lens. If all the galaxies with $m\simgt 19$ were background to the cluster, this would result in an anti-correlation between cluster mass and these galaxies. However, most of them are not. To check the redshift distribution of galaxies, we plot in Fig.~\ref{fig:m_z}, the BPZ redshift estimates of the Subaru galaxy catalogue, for both R-band (blue dots) and Z-band (red crosses) galaxies. The dashed line at the cluster redshift $z=0.308$ is provided for reference.

Using a conservative redshift range for cluster member galaxies, $\Delta z\sim \pm 0.2$, we see that at $m\simlt 22$ most, if not all galaxies are consistent with being in the cluster, while at $m\simgt 22$, a substantial fraction of them is background to the cluster, and should be anti-correlated with the galaxy mass due to magnification bias. Furthermore, faint galaxy counts are likely affected by light from nearby (in projection) brighter galaxies, further contributing to the observed anti-correlations between faint galaxies and mass, because the latter is strongly correlated with bright galaxies. A careful separation of the effect of light contamination and magnification bias is beyond the scope of this paper. Both of these effects result in anti-correlations being stronger for fainter galaxies. We choose two magnitude ranges to illustrate this, $24 \le m \le 25.5$ and $22 \le m \le 24$ (for reference, the $m=24$ line is indicated in Fig.~\ref{fig:m_z}). Fig.~\ref{fig:corr_sub} shows that anti-correlations are observed in both of these ranges, and for both HFFv3 and HFFv4. (Subaru Z-band shows similar results.)

\begin{figure}                
  \includegraphics[width=\columnwidth]{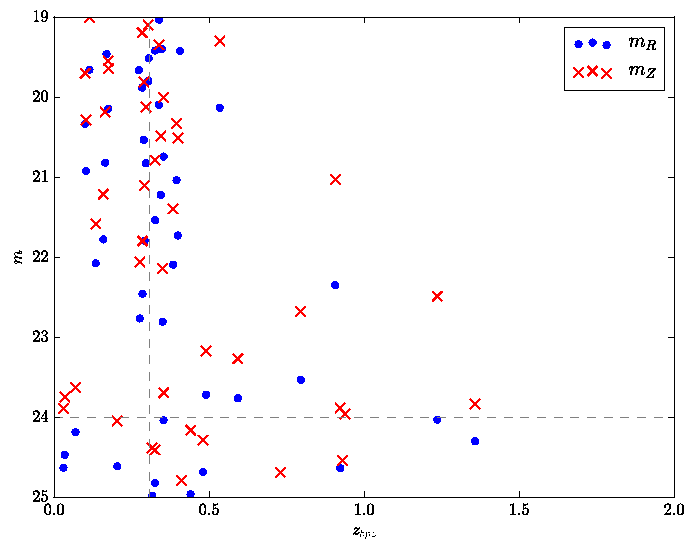}
  \caption{Subaru galaxy BPZ redshifts against their apparent magnitudes. Blue dots indicate R-band Subaru galaxies and red crosses are Z-band Subaru galaxies, in the direction of Abell 2744. Two lines are provided for clarity, one, at the cluster redshift, $z=0.308$, and the other at the magnitude $m=24$ separating two of our magnitude ranges.}\label{fig:m_z}
\end{figure}

\section{Conclusions}
\label{sec:conclusion}

Using exquisite lens image data provided by the Hubble Frontier Field observations, and relying on the collaborative nature of the HFF project, we are able to address two questions pertaining to galaxy clusters. The first question asks how trustworthy, or robust are the uncertainties in the derived mass distribution of galaxy clusters?  The second question is the same that we asked in the analysis of another HFF cluster, MACS J0416 \citep{seb16}; how well does light follow mass in a merging and highly disturbed galaxy cluster? Our genetic algorithm-based lens inversion method, \grale, is free-form, and therefore is the ideal choice for the task, because no information about the cluster or its galaxy members is part of the input. 

We performed two mass reconstructions of Abell 2744. The first reconstruction, HFFv3, used 55 images as inputs with a mix of spectroscopic and photometric redshifts from \cite{ric14,joh14,wan15,jau15}. Our second reconstruction, HFFv4, made use of 91 images with spectroscopic redshifts from \cite{mah18}. The two data sets are reasonably different: only 10 of the 91 images used in HFFv4 have the same positions and redshifts as in HFFv3. 

Both averaged mass maps have small lens-plane RMS's: $0.53$\arcsecond~for HFFv3 and $0.87$\arcsecond~for HFFv4. The overall elongated shape of Abell 2744 and the two major cluster components are clearly seen in both versions. On scales of $6-7$\arcsec, or $\sim 30\,$kpc, there are small, non-statistically significant differences in the concentration of the mass peaks associated with galaxies, that arise primarily due to the differences in the inversion procedure in the two versions. 

To gauge the robustness of \grale~reconstructions against changes in data, as well as small modifications in the inversion procedure, we compared the mass maps of HFFv3 and HFFv4 to each other.  Using \grale's uncertainties we showed that the two mass maps are very similar over most of the cluster area defined by the multiple images. Only about 1.5\% of the area had deviations between the two maps that exceeded $4\sigma$. Similar conclusions about \grale~uncertainties apply to the two reconstructions of Abell 3827, carried out elsewhere. This shows that \grale~derived uncertainties are robust. 

By examining the distribution of local mass peaks around bright galaxies in 40 individual \grale~reconstructions, and by constructing galaxy-mass correlation functions, we find that, in a statistical sense, light follows mass, with brighter galaxies showing stronger correlations than faint ones. This result is seen across both reconstruction versions. We also show that faint galaxies are anti-correlated with mass, with the effect being stronger for fainter galaxies. The anti-correlation is likely the result of two unrelated causes: light contamination from bright galaxies, and lensing magnification bias acting on galaxies background to the cluster.

\section*{Acknowledgments}

The authors are grateful to the Hubble Frontier Field community for generously sharing the multiple image data, often before publication. KS and LLRW acknowledge the computational resources and support of the Minnesota Supercomputing Institute.  
The authors would like to gratefully acknowledge Mario Nonino for providing galaxy data on Abell 2744.



\bsp	
\label{lastpage}
\end{document}